\documentclass[aps,prl,twocolumn,showpacs,letterpaper,superscriptaddress]{revtex4-1}

\usepackage[colorlinks=true, citecolor=blue, linkcolor=blue, urlcolor=blue]{hyperref}

\usepackage{graphicx,dcolumn,longtable,epsfig}
\usepackage[usenames]{color}
\usepackage{amssymb}
\usepackage{amsmath}
\usepackage{epstopdf}
\usepackage{bm}
\usepackage{footnote}
\usepackage{float}
\usepackage{subfigure}
\usepackage{color}

\input epsf.tex

\def\bea{\begin{eqnarray}}
\def\eea{\end{eqnarray}}
\def\be{\begin{equation}}
\def\ee{\end{equation}}

\begin{document}

\title{Equilibrium phases of two-dimensional bosons in quasi-periodic lattices}
\author{ C. Zhang }
\affiliation{Homer L. Dodge Department of Physics and Astronomy,
The University of Oklahoma, Norman, Oklahoma ,73019, USA}
\author{ A.~Safavi-Naini}
\affiliation{Homer L. Dodge Department of Physics and Astronomy,
The University of Oklahoma, Norman, Oklahoma ,73019, USA}
\affiliation{
JILA , NIST and Department of Physics, University of Colorado, 440 UCB, Boulder, CO 80309, USA}
\author{B. Capogrosso-Sansone}
\affiliation{Homer L. Dodge Department of Physics and Astronomy,
The University of Oklahoma, Norman, Oklahoma ,73019, USA}

\begin{abstract}
We report on results of Quantum Monte Carlo simulations for bosons in a two dimensional quasi-periodic optical lattice. We study the ground state phase diagram at unity filling and confirm the existence of three phases: superfluid, Mott insulator, and Bose glass. At lower interaction strength, we find that sizable disorder strength is needed in order to destroy superfluidity in favor of the Bose glass. On the other hand,  at large enough interaction superfluidity is completely destroyed in favor of the Mott insulator (at lower disorder strength) or the Bose glass (at larger disorder strength). At intermediate interactions, the system undergoes an insulator to superfluid transition upon increasing the disorder, while a further increase of disorder strength drives the superfluid to Bose glass phase transition.
While we are not able to discern between the Mott insulator and the Bose glass at intermediate interactions, we study the transition between these two phases at larger interaction strength and, unlike what reported in~\cite{Sandvik2014} for random disorder, find no evidence of a Mott-glass-like behavior.
\end{abstract}

\maketitle

{\em{Introduction:}} Condensed matter systems, either manufactured or occurring in nature, posses, in general, a certain degree of disorder. Studying physical phenomena such as Anderson~\cite{Anderson} localization, resulting from the presence of disorder, is therefore of crucial importance.  
Anderson localization pertains to the case of non-interacting fermions. More realistic systems, though, consist of interacting particles. For interacting systems, the interplay between disorder and interaction may result in novel physical effects. For instance, when random disorder is added to paradigmatic condensed matter models, such as the Bose-Hubbard model or the BCS model for superconductivity, it gives rise to disorder-driven phase transitions from a conducting to an insulating phase, resulting from the localization of bosons and cooper pairs, respectively~\cite{Fisher1,Pollet:2013cp,Ghosal:1998ja,
Dubi:2007kz}. While disorder driven phase transitions have been observed in a wide range of experimental systems such as films of adsorbed $^{4}$He on substrates \cite{He4,Ebey:1998jt}, bosonic magnets~\cite{Tao2010,Rong2012,Rong2012Nature}, and thin superconducting films \cite{ ThinSuperconductingFilms,Schneider:2012id}, and in spite of a remarkable theoretical effort \cite{Fisher:1990TF,Fisher:1990SC,Reppy:1992il,Moon:1995ds, Magnetic3, Nohadani:2005bua,Yamada:2011gs,Wulf:2013eva}, 
a thorough understanding of the effects of disorder in interacting quantum many body systems is lacking. On the one hand these systems are challenging to study theoretically, on the other poor control over experimental condensed matter systems does not allow for thorough experimental investigation of these systems. 

Optical lattice systems of ultra-cold atoms and molecules provide a unique possibility of engineering matter with an unprecedented level of control and flexibility over the parameters entering the hamiltonian\cite{Bloch:2005gn,Morsch:2006em,ColdAtom1,ColdAtom2}. 
Hence, optical lattice simulators have rapidly become an important tool in the study of disordered systems where disordered or quasi-disordered optical lattice potentials are created using speckle patterns or multi-chromatic incommensurate optical lattices respectively~\cite{CreatDisorder1,CreatDisorder2,DeMarco2009PRL,AAspect2012}. These techniques were employed in the first realizations of Anderson localization in one dimension using non-interacting bosons in continuum~\cite{AAspect:2008kd}, and in an optical lattice~\cite{Inguscio2008}. 
Subsequently, delocalization induced by weak repulsive interaction was observed in one dimension~\cite{Inguscio2010}, while localization induced by strong interaction was demonstrated in one and three dimensions~\cite{Schneble2011,DeMarco2009PRL,DeMarco2010Nature}. 

In the past decade the behavior of strongly interacting systems in the presence of random disorder has been studied extensively using a variety of theoretical methods~\cite{Pollet:2013cp,Ceperley1991,Rapsch:1999bw,ScalingFunction2,Pollet2009BG,Pollet2009PRB,Soyler:2011ik,Niederle:2013jy,Schmitteckert:1998ee,Ceperley2011}.
For the most part these studies have considered systems of bosonic particles trapped in one-, two-, or three-dimensional optical lattices. In the absence of disorder, these systems feature two phases: superfluid (SF) and Mott-Insulator (MI).
In the presence of random disorder a third insulating but compressible phase, known as the Bose Glass (BG), is stabilized \cite{Fisher1}. As a result of finite disorder strength, no direct SF-MI phase transition exists~\cite{Pollet2009BG} and the BG always intervenes between the MI and SF regions.

Optical lattice systems can also be used to create quasi-periodic trapping potentials by employing bichromatic lattices which are formed by combining two optical lattices with incommensurate wavelengths \cite{CreatDisorder2}. In solid state physics, quasi-periodic crystalline structures such as photonics quasicrystals were found to have a nontrivial connection to topological states of matter \cite{Kraus:2012iq,Kraus:2012kz,Verbin:2013bv}. Non-intereacting bosons in quasi-periodic one-dimensional potentials are described by the analytically solvable Aubry-Andr\'{e} model \cite{AAOriginal,Modugno2009}, which features Anderson localization. This model was first realized experimentally in \cite{Inguscio2008}, where Anderson localization was confirmed by investigating transport properties, as well as the spatial and momentum distributions.
As in the case of random disorder, the introduction of interactions to the Aubry-Andr\'{e} model increases the complexity of the system and gives rise to new physical phenomena. Most of the recent theoretical work has focused on one-dimensional systems of interacting bosons in quasi-periodic potentials, where DMRG methods can be successfully employed \cite{Louis:2007ds,Giamarchi2008,Minguzzi2009,Modugno2009PRA,Deng2008,Cai:2011eq}. These studies have identified a direct SF-MI transition of Kosterlitz-Thouless type at weak disorder~\cite{Deng2008,Giamarchi2008}, in accordance to the predictions of the Harris-Luck criterion\cite{Harris-Luck}. The criterion states that a perturbative quasi-periodic disorder is \emph {irrelevant} from the renormalization group point of view, leaving the nature of the phase transition unchanged when compared to the transition in the clean system.

In the following, we use Path Integral Quantum Monte Carlo by the Worm algorithm \cite{Prokofev:1998gz} to study two-dimensional (2D) lattice bosons in the presence of quasi-periodic disorder. We find that if the interaction strength is smaller than the critical interaction strength corresponding to the 2D SF-MI transition in the clean system, sizable disorder is needed to destroy superfluidity. On the other hand, at any given disorder, one can find an interaction strength above which superfluidity is completely destroyed in favor of an insulating phase. At lower disorder strength this insulating phase is a MI, while at larger disorder strength it is a BG. Our numerical results for the compressibility in the range of interaction strengths where SF has completely disappeared are consistent with a direct MI-BG phase transition and do not show any evidence of a cross-over region characterized by Mott-glass-like behavior (or anomalous Bose glass), unlike the findings of~\cite{Sandvik2014} for the case of random disorder. Finally, at intermediate interaction strengths, the system undergoes an insulator to superfluid transition upon increasing the strength of the disorder. One can (re)enter the Bose glass phase by further increasing of disorder strength.

{\em{Hamiltonian:}} We study a system of bosons in a 2D lattice in the presence of quasi-periodic disorder, described by the Hamiltonian
\begin{align}
\label{eq:H}
\nonumber H&=-J\sum_{\langle i\,j\rangle }(a_i^\dagger a_j+h.c)+\frac{U}{2}\sum_{i}n_i(n_i-1)\\
&-\mu \sum_i n_i+ \sum_i \Delta_i n_i.
\end{align}
The first term in the Hamiltonian is the kinetic energy, where $a_i^\dagger$ ($a_i$) are the bosonic creation (annihilation) operators with the usual commutation relations, and $J$ is the hopping matrix element between sites $i$ and $j$. We use $\langle \dots \rangle$ to denote nearest neighboring sites. Here $U$ sets the strength of the on-site repulsion and $\mu$ is the chemical potential, which in the absence of disorder, sets the number of particles in the system. The quasi-periodic on-site disorder $\Delta_i$ is created by perturbing the primary optical lattice with a second incommensurate one. The net result is a quasi-periodic external potential that couples to the on-site density $n_i$. Hence, the on-site disorder takes the form $\Delta_i =\Delta \cos (2 \pi \beta_d x_i + \phi_x)\cos (2 \pi \beta_d y_i + \phi_y)$, where $\Delta$ is the strength of disorder, $\phi_{x,y}$ is an arbitrary phase shift, and $\beta_d$ measures the degree of commensurability. Both $\Delta$ and $\beta_d$ can be tuned experimentally, the former by tuning the relative heights of the primary and secondary lattices and the latter by varying the wave numbers of the two lattice potentials. The results presented below correspond to the maximally incommensurate ratio given by the choice $\beta_d=(\sqrt5-1)/ 2$.

{\em{Results:}} In the following we present a numerical study of the Hamiltonian \eqref{eq:H} at unit filling ($n=N/N_{\rm sites}=1$) by the means of quantum Monte Carlo simulations using the Worm Algorithm. In order to obtain accurate results in the thermodynamic limit we perform finite-size scaling on the simulations results. This process is challenging in the presence of quasi-periodic disorder where the disorder is incommensurate with the lattice. Incommensurability means that one cannot produce comparable systems by simply scaling the lattice size. To circumvent this problem we have used system sizes $L$, with $N_{\rm site}=L\times L$, from the Fibonacci sequence~\cite{Santos}. Unlike the results reported for one-dimensional systems~\cite{Santos}, we have found that for disorder strength $\Delta \gtrsim 3J$ our results depend strongly on the choice of $(\phi_x,\phi_y)$. Hence, for each set of parameters $(\mu, \Delta, U, L)$ we have run simulations with fifty different choices for phases $\phi_{x,y}\, \in \, \left[0,\, 2\pi\right)$. The results presented below are extracted from the fifty runs using the bootstrap method. We find that further averaging over $(\phi_x, \phi_y)$ realizations simply reduces the statistical error. 

\begin{figure}[h]
\includegraphics[width=0.5\textwidth]{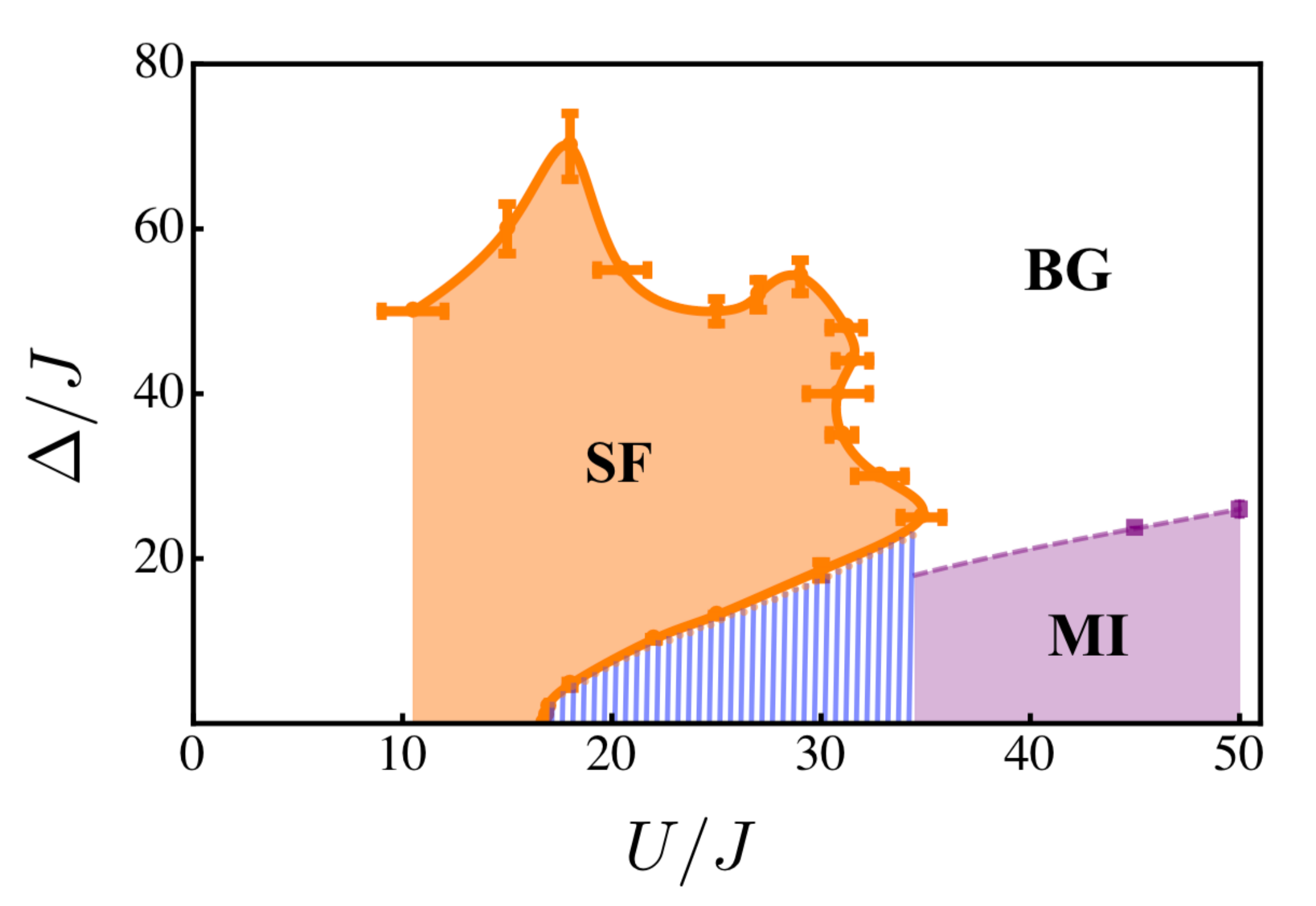}
\caption{(Color online) Ground state phase diagram of the system described by Eq.~\ref{eq:H} at filling factor $n=1$. The horizontal and vertical axes are the onsite interaction strength $U/J$ and disorder strength $\Delta/J$, respectively. Using these two parameters as tuning knobs, the system can form a Mott-Insulator (MI), a superfluid (SF), and a Bose glass (BG). Simulations results for the SF-insulator phase boundary are shown using solid orange circles (the solid orange line is a guide to the eye), while solid purple squares (the dashed line is a guide to the eye) correspond to the phase boundary between the MI and BG phases. At lower disorder  and intermediate interactions we are unable to distinguish between the MI and the BG (dashed blue region). }
\label{fig:t0pd}
\end{figure}

The ground state phase diagram of the system at unit filling is shown in Fig.~\ref{fig:t0pd}, where the horizontal and vertical axes correspond to $\frac U J$ and $\frac \Delta J$ respectively. At lower disorder strength and for $U/J\lesssim 16$ the system is in the superfluid state associated with the presence of off-diagonal long-range order. The superfluid phase is characterized by finite compressibility and non-zero single particle condensate order parameter, $\langle \psi \rangle=\langle a_i\rangle \neq 0$, associated with a finite superfluid stiffness $\rho_s$. The superfluid stiffness is extracted from simulations using the relation $\rho_s=\langle \mathbf W^2 \rangle /dL^{d-2}\beta$, where $\mathbf W$ is the winding number in space, $d$ is the spatial dimension ($d=2$ in our case)\cite{Winding}, and $\beta=1/T$ is the inverse temperature. In all our simulations we have chosen $\beta$ such that the system is in its ground state, and have scaled $\beta\propto L^z$ where $z$ is the dynamical critical exponent. The SF phase becomes unstable at stronger disorder strength and a transition to the insulating BG phase occurs. The BG phase is characterized by vanishing superfluid stiffness and finite compressibility $\kappa$.

For $16\lesssim U/J\lesssim 35$ and at low disorder strength, the system is in an insulating phase and undergoes a phase transition in favor of the SF phase upon increasing the disorder strength. A similar phase transition is present if the trapping potential features random disorder, where it has been shown that the presence of an intervening BG phase between the MI and SF is guaranteed by the theorem of inclusions~\cite{Pollet2009BG}. It should be noted that this theorem does not apply to quasi-periodic disorder and therefore the existence of the BG phase or the lack thereof should be confirmed by direct measurement of the compressibility. However, the parameter regime corresponding to the range of interactions and disorder strengths where the BG region may form is narrow.  As for the case of random disorder, the compressibility of the BG in narrow regions would be too small to be detected numerically, making it impossible to distinguish between the MI and BG phases. We are therefore unable to discuss the onset of the BG phase, and generically refer to the dashed blue region separating the SF and the zero-disorder MI in Fig.~\ref{fig:t0pd} as an insulating phase. Further increasing the disorder strength results in the destruction of the SF order in favor of the BG.  

\begin{figure}[h]
\includegraphics[width=0.5\textwidth]{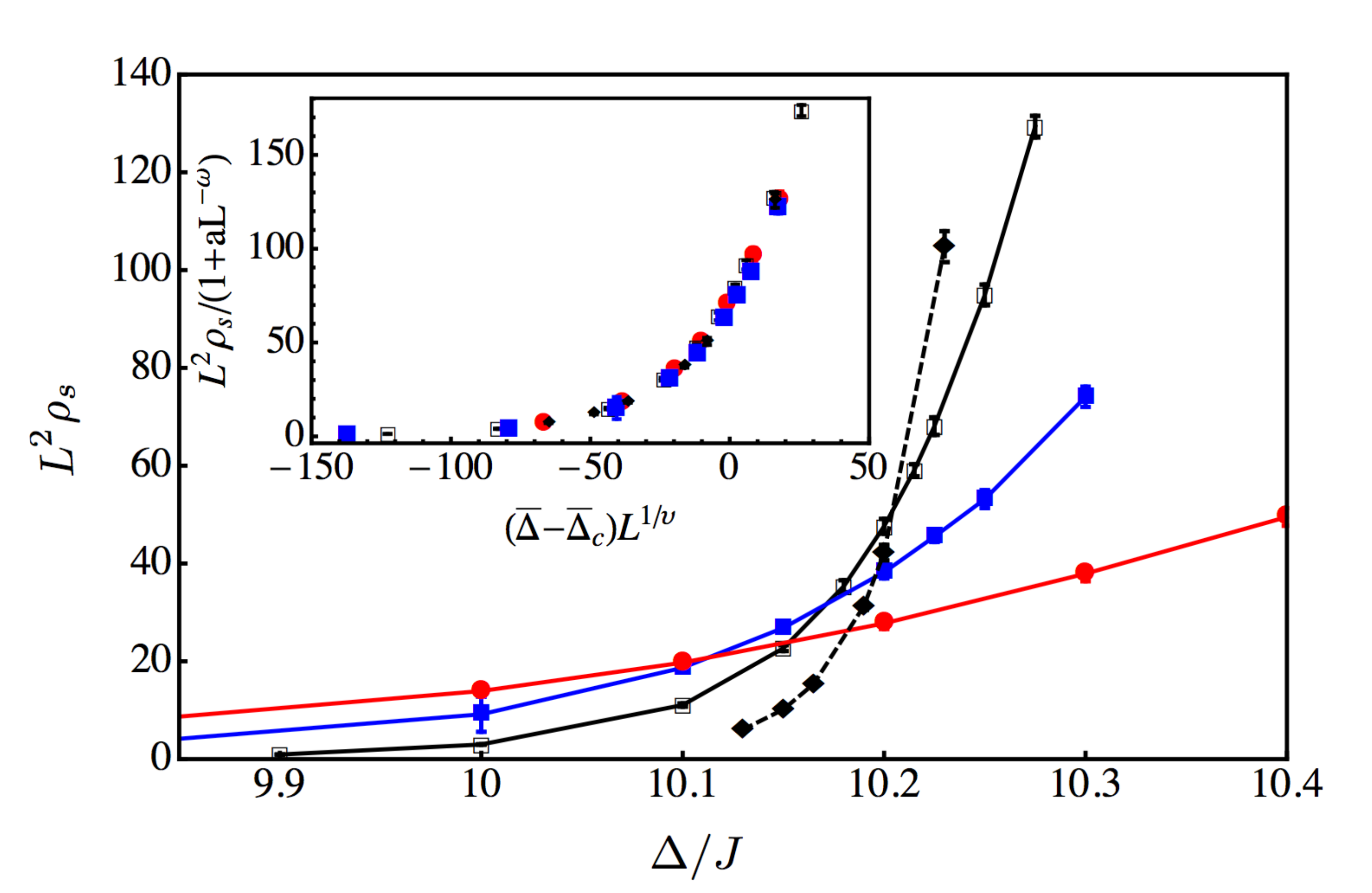}
\caption{(Color online) Main plot: Scaled superfluid stiffness $\rho_sL^{-(d+z-2)}$ with $z=2$, as a function of $\Delta/J$ for $U/J=22$ and $L=$ 21, 34, 55 and 89 using red circles, blue squares, empty black squares, and black diamonds, respectively. In these simulations we have used $\beta=(L/2)^2$ to scale the imaginary time dimension $L_\tau$. Inset: Data collapse using $\nu=0.67$, $a=-9.4$, $\omega=-0.9$, and $\bar \Delta_c=10.21$ corresponding to the critical point extracted from the main plot. Here $\bar \Delta=\Delta/J$. The symbols are the same as those used for the main plot.  }
\label{fig:U22cross}
\end{figure}

Figure~\ref{fig:U22cross} illustrates the finite size scaling procedure used to determine the SF-BG (or generic insulator) phase boundary (solid orange circles in Fig.~\ref{fig:t0pd}). Here we plot the scaled superfluid stiffness $\rho_sL^{(d+z-2)}$ with $z=2$, as a function of $\Delta/J$ at $U/J=22$ and $L=$ 21, 34, 55, and 89 (red circles, blue squares, empty black squares, and black diamonds, respectively). In these simulations we have used $\beta=(L/2)^z$ to scale the imaginary time dimension $L_\tau$. The dynamical critical exponent $z$ was set to $d=2$, following the prediction in Ref. \cite{Fisher1} for random disorder, and the recent unambiguous confirmation using Monte Carlo techniques \cite{NikolayPsi}. The drift in the position of the intersection point indicates that a correction to the finite size scaling relation $\rho_sL^{(d+z-2)}=f(L^{1/\nu}\frac{\Delta}{J},\beta L^{-z})$ where $f(x,\rm {const})$ is a universal function, must be included in order to observe data collapse. After this correction is taken into account the scaling relation takes the form $\rho_sL^{(d+z-2)}=\left(1+aL^{-\omega}\right) f(L^{1/\nu}\frac{\Delta}{J},\beta L^{-z})$\cite{Correction}. The inset of Fig.~\ref{fig:U22cross} shows $L^2 \rho_s/\left(1+aL^{-\omega}\right)$ as a function of $(\bar \Delta-\bar\Delta_c)L^{1/\nu}$, where $\bar \Delta=\Delta/J$.  From the best data collapse we find $\nu=0.67\pm 0.07$ ($\nu=0.67$ holds for the SF-insulator transition of a clean system), $a=-9.4\pm0.5$, $\omega=-0.9\pm 0.05$, and $\bar \Delta_c=10.21\pm0.05$. This value of $\nu$ suggests that the quasi-periodic disorder is still irrelevant for $\Delta/J\sim 10$.
As noted above, the choice of dynamical exponent $z=d$ has only been predicted and confirmed for random disorder. To ensure that our choice of the critical exponent $z$ does not affect the position of the transition line, we have performed finite size scaling with a choice of $z=1.5$ for various points on the transition line. We find that the critical point remains the same within the error bars for the two different choices of $z$. 

\begin{figure}[h]
\includegraphics[width=0.48\textwidth]{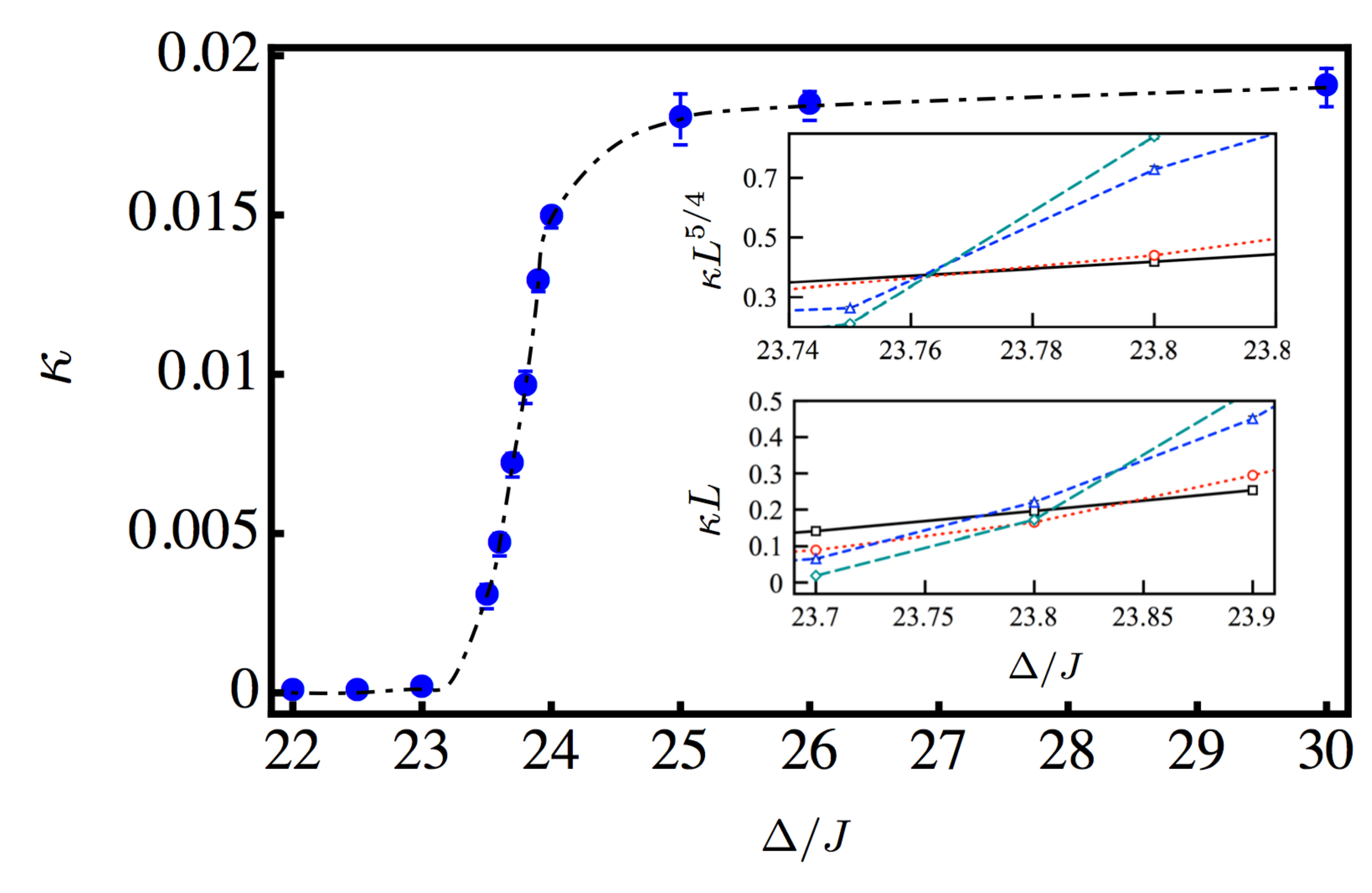}
\caption{(Color online) Main plot: $\kappa$ vs. $\Delta/J$ for $U/J=45$, $L=21$, and $\beta=L/2$. The compressibility becomes finite at $\Delta/J\sim 23.5$ and plateaues at $\Delta/J\sim 24.5$. Inset: The top and bottom panels show $\kappa L^{d-z}$ versus $\Delta/J$ for $z=0.75$ and $z=1$ respectively, at $U/J=45$. The scaled compressibility is shown for $L=21,$ 34, 55, and 89 using black squares,  red circles, blue triangles and green diamonds, respectively. Our data indicates that $0.75\lesssim z\lesssim 1.25$. }
\label{fig:compcross}
\end{figure}

Finally, we turn our attention to the region $U/J\gtrsim35$ where the SF phase is completely absent and the system undergoes a MI-BG transition upon increasing the disorder strength. In this region of the parameter space we can easily measure compressibility and distinguish between the MI and BG phases. To this end we have performed simulations with $z=$0.75, 1, 1.25, and 1.5 at $U/J=45$.  The inset of Fig.~\ref{fig:compcross} shows the scaled compressibility $\kappa L^{d-z}$ with $z=0.75$ (top panel) and $z=1$ (lower panel) for $L=$21, 34, 55, and 89 using black squares,  red circles, blue triangles and green diamonds, respectively. While we have not performed an exhaustive scan over different values of dynamical exponent $z$, we find that the best crossing corresponds to $z=0.75$ and gives the critical point at $\Delta_c/J=23.76\pm 0.05$.

\begin{figure}[h]
\includegraphics[width=0.48\textwidth]{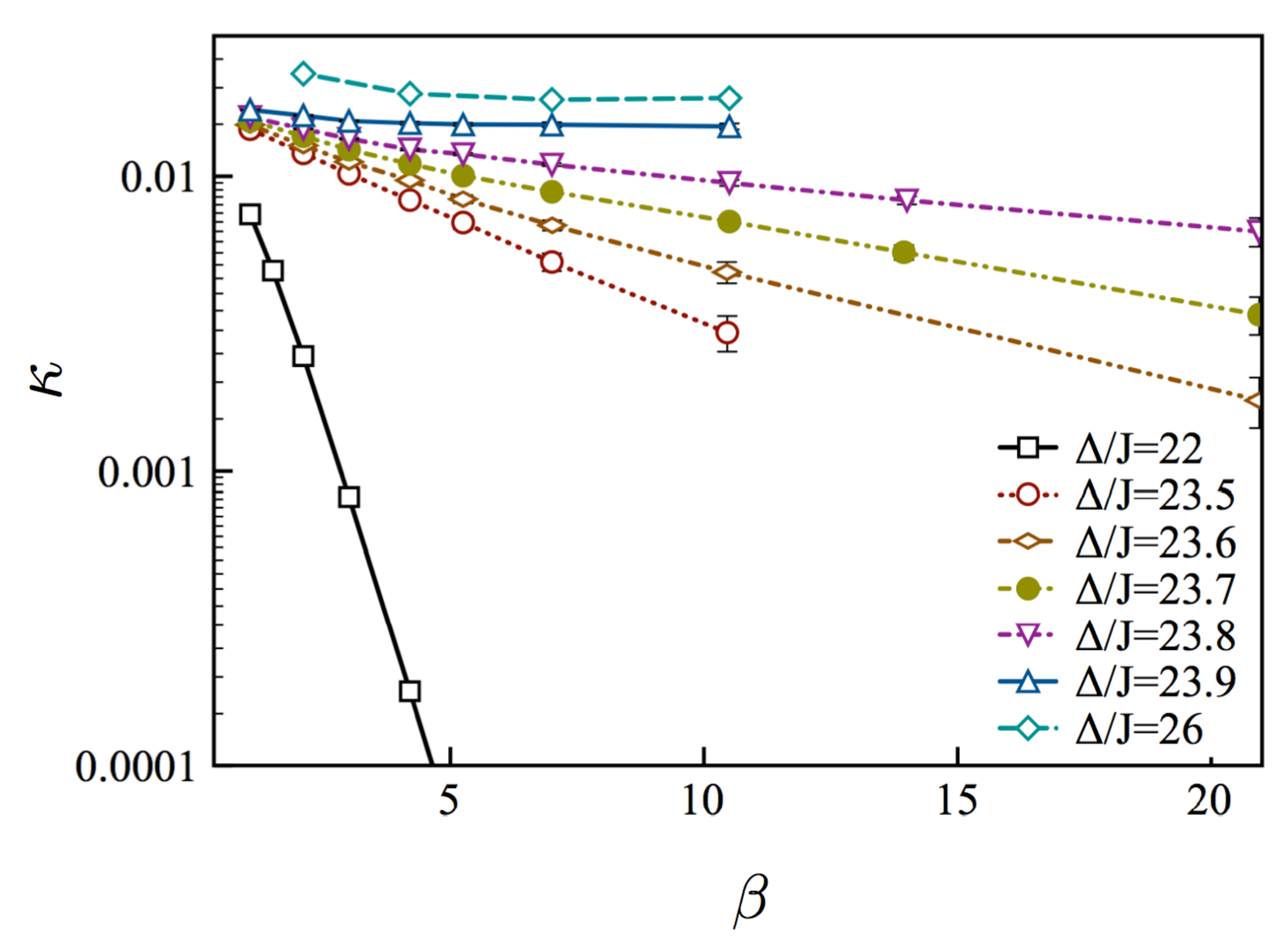}
\caption{(Color online) The plot shows $\log \kappa$ as a function of $\beta$, at $U/J=45$ and $L=21$ for $\Delta/J=$22, 23.5, 23.6, 23.7, 23.8, 23.9, and 26. Below the critical point $\Delta/J=23.76\pm 0.05$ (see inset of Fig.~\ref{fig:compcross}) the behavior is consistent with that of a MI. Above the transition, the behavior is consistent with that of the BG phase.
}
\label{fig:compbeta}
\end{figure}

Lastly, we discuss the possibility of the existence of a cross-over region separating the MI from the BG, where the system forms a Mott glass or possesses a Mott-glass-like anomalous BG behavior as discussed in~\cite{Sandvik2014} for the case of random disorder. A Mott glass is a gapless yet incompressible insulator. In \cite{Sandvik2014} the authors present numerical results which suggest that there exists a region in parameter space where the system possesses Mott-glass-like behavior, i.e. negligible compressibility $\kappa$, with $\kappa\sim \exp(-b/T^\alpha)+c$ ($\alpha<1$ and $c\sim0$).
In analogy with~\cite{Sandvik2014}, we have studied the behavior of $\kappa$ away from the SF lobe boundary at fixed $\beta$ as a function of $\Delta/J$, and at fixed $\Delta/J$ as a function of $\beta$. The main plot of figure~\ref{fig:compcross} shows $\kappa$ vs. $\Delta/J$ for $U/J=45$, $L=21$, and $\beta=L/2$. The compressibility becomes finite at $\Delta/J\sim 23.5$ and plateaus at $\Delta/J\sim 24.5$. 
Fig.~\ref{fig:compbeta} shows $\kappa$ as a function of $\beta$ at $U/J=45$ and $L=21$ for $\Delta/J=$22, 23.5, 23.6, 23.7, 23.8, 23.9, and 26. Our data indicates that below the quantum critical point, $\Delta_c/J=23.76\pm 0.05$, the system is in the MI state and $\kappa\sim \exp(-\beta \Delta_G)$, where $\Delta_G$ is the energy gap. Upon increasing the disorder strength the system enters the BG phase as shown by a plateaued compressibility at large enough $\beta$ (see curves corresponding to $\Delta/J=$23.9 and 26). It should be noted that at $\Delta/J=23.8$ we observe MI-like behavior which can be attributed to the finite size of the system. The numerical results shown in Figure~\ref{fig:compcross} and~\ref{fig:compbeta} strongly support the absence of a cross-over region where the system behaves like a Mott glass. If this crossover region exists at $U/J=45$, Figure~\ref{fig:compbeta} suggests that it would only extend within a range of disorder strength of width $\sim 1\%$.

In conclusion, we have used Path Integral Quantum Monte Carlo by the Worm algorithm to study the phase diagram of bosons in a two-dimensional quasi-periodic optical lattice. As in the case of random disorder, the ground state phase diagram contains three phases: superfluid, Mott insulator, and Bose glass. At weaker interactions, the superfluid phase is favored and significant disorder has to be introduced in order to destroy superfluidity. At strong enough interactions, the superfluid phase has disappeared, and for weak enough disorder the system forms a Mott insulator. Upon increasing the disorder strength the system undergoes a phase transition from Mott insulator to Bose glass. We have used finite temperature simulations to establish that there is no Mott-glass-like Bose glass behavior separating the Mott insulator from the Bose glass. Finally, at intermediate interaction strengths and lower disorder, the compressibility of the Bose glass is too small to be measured numerically in finite systems. In this region we are unable to distinguish between a Mott insulator and a Bose glass.

\emph{Acknowledgments} We would like thank N. Prokof'ev and \c S.G. S\"oyler for useful discussions. This work was supported by the National Science Foundation (NSF) under the grant NSF-PIF-1415561. ASN was supported by AFOSR-MURI grant. The computing for this project was performed at the OU Supercomputing Center for Education $\&$ Research (OSCER) at the University of Oklahoma (OU). 

\bibliography{quasi}


\end{document}